\documentclass[11pt]{article}
\usepackage{amsfonts,amsmath}
\usepackage{graphicx}

\def\bbt{\bibitem}
\def\be{\begin{equation}}
\def\en{\end{equation}}
\def\ber{\begin{eqnarray}}
\def\enr{\end{eqnarray}}
\def\nmb{ \nonumber\\}
\def\d{\partial}
\def\rbr{\rbrack}
\def\lbr{\lbrack}
\def\rbrc{\rbrace}
\def\lbrc{\lbrace}
\def\lgl{\langle}
\def\rgl{\rangle}
\def\ov{\over }

\def\al{\alpha}

\def\dlt{\delta}

\begin{document}

%\nopagenumbers
\rightline{Landau Tmp/08/2014.}
%\rightline{August, 2014}

\vskip 2 true cm
\centerline{\bf Kazama-Suzuki Models of N=2 Superconformal Field Theory and Manin triples.}
\vskip 20pt 
\centerline{S. E. Parkhomenko}

%\footnote{$\sp{\dag}$}
% {e-mail address:spark@itp.ac.ru}}

\vskip 1 true cm
\centerline{\bf Abstract}
\vskip 0.5 true cm

 Kazama-Suzuki coset models is an interesting class of N=2 supersymmetric models of conformal field theory which are used to build realistic models of superstring in 4 dimensions. We formulate Kazama-Suzuki construction of N=2 superconformal coset models using more general language of Manin triples and represent the corresponding N=2 Virasoro superalgebra currents in explicit form. A correspondence between the Kazama-Suzuki models and Poisson homogeneous spaces is established also.

\smallskip
\vskip 10pt

 It is by now well-known due to Gepner \cite{Gep} that the unitary N=2 superconformal field theories play an important role in the construction of realistic models of superstring compactification from 10 to 4 dimensions. The idea of his construction is to use N=2 superconformal field theories with central charge 9 for the internal sector of the string degrees of freedom. In particular, Gepner considered a product of N=2 minimal models such that their total central charges adds up to 9. Then he showed that GSO projection can be obtained by selecting the states with odd integral $U(1)$ charges and this projection can be made consistent with modular invariance if one adds at the same time the spectral flow twisted sectors (see \cite{Gep}). 
 
 In the paper \cite{KS} Kazama and Suzuki constructed a new large class of N=2 superconformal field theories, including N=2 superconformal minimal models, using the coset space method \cite{GKO}. 
It rised the problem of generalization of the Gepner's construction to include these new models and investigate the corresponding models of the superstring compactification from 10 to 4 dimensions. However this problem is quite far from its solution untill now mainly because of the Kazama-Suzuki models themselves has not yet been studied quite well.

 As a particular case of Kazama-Suzuki models N=2 superconformal WZNW models was investigated in \cite{SSTP} and \cite{P} (see also \cite{Getz}) where a one to one correspondence between the Manin triples, which are the classical limit of Drinfeld's quantum dubles \cite{Drin} and N=2 superconformal WZNW models was established. Since then it has become clear that this important relation should also takes place for N=2 superconformal coset models and Kazama-Suzuki construction can be formulated naturally in terms of Manin triples also. However by some of the reasons the explicit formulas has not been represented in the literature in spite of its importance at least in the context of Poisson-Lie T-duality \cite{K}, \cite{P1}, \cite{KP} of the superstring vaccua. 

 In this note we fill up this defect formulating Kazama-Suzuki construction on the language of Manin triples and represent the corresponding N=2 Virasoro superalgebra currents entirely in terms of Manin triple and Manin subtriple objects. It allows us to establish a correspondence between the Kazama-Suzuki models and Poisson homogeneous spaces. 

 We begin with the definition of Manin triple \cite{Drin}. 

{\bf Definition.} A Manin triple $(g,g_{+},g_{-})$ consists of
a Lie algebra $g$, with nondegenerate invariant inner product
$(,)$ and isotropic Lie subalgebras $g_{\pm}$ such that
$g=g_{+}\oplus g_{-}$ as a vector space.

For any finite dimensional Manin triple let us fix arbitrary
orthonormal basis $\{E^{A}, E_{A}$, $A= 1,...,D\}$ in algebra
$g$ so that $\{E^{A}\}$ is a basis in $g_{-}$, $\{E_{A}\}$
is a basis in $g_{+}$. The commutators and Jacobi identity of $g$
have the form
\ber  \lbr E^{A},E^{B}\rbr=f^{AB}_{C}E^{C},     \nmb
           \lbr E_{A},E_{B}\rbr=f_{AB}^{C}E_{C},     \nmb
           \lbr E^{A},E_{B}\rbr=f_{BC}^{A}E^{C}-f^{AC}_{B}E_{C}, 
           \label{D.1}
\enr

\ber f^{AB}_{D}f^{DC}_{E}+f^{BC}_{D}f^{DA}_{E}+f^{CA}_{D}f^{DB}_{E}=0, 
\nmb
f_{AB}^{D}f_{DC}^{E}+f_{BC}^{D}f_{DA}^{E}+f_{CA}^{D}f_{DB}^{E}=0,                                         
\nmb
f_{MC}^{A}f^{BM}_{D}-f_{MD}^{A}f^{BM}_{C}-f_{MC}^{B}f^{AM}_{D}+ f_{MD}^{B}f^{AM}_{C}= 
f_{CD}^{M}f^{AB}_{M}. 
\label{D.2}
\enr

In what follows the following consequence of (\ref{D.2})
\ber
f_{M}f^{MB}_{A}+f^{M}f_{MA}^{B}=f_{NM}^{B}f^{NM}_{A},  
\label{D.3}
\enr
where $f_{M}=f_{MA}^{A}$, $f^{M}=f^{MA}_{A}$, will be important to us.

 Denote by $\langle,\rangle$ the Killing form of $g$. It is not difficult
to find
\ber 
\langle E^{A}, E^{B}\rangle= 2f^{AC}_{D}f^{BD}_{C}, 
\nmb
\langle E_{A}, E_{B}\rangle= 2f_{AC}^{D}f_{BD}^{C}, 
\nmb
\langle E^{A}, E_{B}\rangle= -f^{CD}_{B}f_{CD}^{A}-2f^{AC}_{D}f_{BC}^{D}, 
\label{D.4}
\enr
and rewrite the last expression from (\ref{D.3}) in the form
\ber
 \langle E^{B},E_{C}\rangle= -K^{B}_{C}-2A^{B}_{C} 
 \label{D.5}
\enr
where
\ber 
K^{B}_{A}= f_{C}f^{CB}_{A}+f^{C}f_{CA}^{B}, 
\nmb
A^{B}_{C}= f_{CM}^{D}f^{BM}_{D}. 
\label{D.6}
\enr

Let $J^{A}(z), J_{A}(z)$ be the generators of the affine Kac-Moody
algebra $\hat{g}$, that correspond to the fixed
basis $\{E^{A}, E_{A}\}$, so that currents $J^{A}$ generate
subalgebra $\hat{g}_{-}$ and currents $J_{A}$ generate
subalgebra $\hat{g}_{+}$. The singular OPEs between these
currents is the following

\ber J^{A}(z)J^{B}(w)=-(z-w)^{-2}{1\ov 2}\lgl E^{A},E^{B}\rgl
          +(z-w)^{-1}f^{AB}_{C}J^{C}(w)+reg,   
					\nmb
          J_{A}(z)J_{B}(w)=-(z-w)^{-2}{1\ov 2}\lgl E_{A},E_{B}\rgl
          +(z-w)^{-1}f_{AB}^{C}J_{C}(w)+reg,   
					\nmb
          J^{A}(z)J_{B}(w)=(z-w)^{-2}{1\over 2}
          (q\delta^{A}_{B}-\lgl E^{A},E_{B}\rgl)      
          +(z-w)^{-1}(f_{BC}^{A}J^{C}-f^{AC}_{B}J_{C})(w)+reg, 
\label{D.7}
\enr
where $q$, is a number.

 Let $\psi^{A}(z), \psi_{A}(z)$ be free fermion currents which
have singular OPEs with respect to the inner product $(,)$
\ber
 \psi^{A}(z)\psi_{B}(w)= (z-w)^{-1}\delta^{A}_{B}+reg. 
\label{D.8}
\enr

It has been shown in \cite{P} that the currents

\ber G^{+}={2\ov \sqrt{q}}(\psi^{A}J_{A}-
          {1\over 2}f_{AB}^{C}:\psi^{A}\psi^{B}\psi_{C}:),   
\nmb
G^{-}={2\ov\sqrt{q}}(\psi_{A}J^{A}-
          {1\over 2}f^{AB}_{C}:\psi_{A}\psi_{B}\psi^{C}:),   
\nmb
          K= (\delta^{B}_{A}+{2\ov q}K^{B}_{A}):\psi^{A}\psi_{B}:+
          {2\over q}(f_{C}J^{C}-f^{C}J_{C}),             
\nmb
T={1\over q}:(J^{A}J_{A}+J_{A}J^{A}):+  
          {1\ov 2}:(\d \psi^{A}\psi_{A}-\psi^{A}\d \psi_{A}): 
\label{D.9}
\enr
satisfy the N=2 Virasoro superalgebra OPE's:
\ber
T(z_{1})T(z_{2})=z_{12}^{-4}{c\ov 2}+z_{12}^{-2}2T(z_{2})+z_{12}^{-1}\d T(z_{2})+reg.,
\nmb
T(z_{1})K(z_{2})=z_{12})^{-2}K(z_{2})+z_{12}^{-1}\d K(z_{2})+reg.,
\nmb
T(z_{1})G^{\pm}(z_{2})=z_{12})^{-2}{3\ov 2}G^{\pm}(z_{2})+z_{12}^{-1}\d G^{\pm}(z_{2})+reg.,
\nmb
K(z_{1})K(z_{2})=z_{12}^{-2}{c\ov 3}+reg.,
\nmb
K(z_{1})G^{\pm}(z_{2})=\pm z_{12})^{-1}G^{\pm}(z_{2})+reg.,
\nmb
G^{+}(z_{1})G^{-}(z_{2})=z_{12}^{-3}{2c\ov 3}+z_{12}^{-2}2K(z_{2})+z_{12}^{-1}(2T(z_{2})+\d K(z_{2}))+reg.,
\nmb
G^{+}(z_{1})G^{+}(z_{2})=reg., \ G^{-}(z_{1})G^{-}(z_{2})=reg..
\label{D.Vir}
\enr
where the central charge
\ber
 c= 3(D+{2\ov q}K^{C}_{C}) 
\label{D.10}
\enr
and $z_{12}=z_{1}-z_{2}$.

 Let us fix some Manin subtriple 
\ber
(h,h_{+},h_{-})\subset (g,g_{+},g_{-})
\label{D.11}
\enr
such that $h_{-}$ is spanned by the subset $\lbrc E^{\al}\rbrc$, $\al=1,...,D-d$ of the generators $\lbrc E^{A}\rbrc$, and 
$h_{+}$ is spanned by the subset $\lbrc E^{\al}\rbrc$, $\al=1,...,D-d$ of the generators $\lbrc E_{A}\rbrc$. In what follows we use the small Latin letters to denote the indexes for the remaining generators $E^{a}$ and $E_{a}$ which span $t_{-}=g_{-}/h_{-}$ and 
$t_{+}=g_{+}/h_{+}$ correspondingly.

 The Manin triples formulation of the conditions found by Kazama and Suzuki \cite{KS} is given by the following

{\bf Theorem.} Suppose the subspaces $t_{+}$ and $t_{-}$ are subalgebras. Then the currents
\ber 
G^{+}={2\over \sqrt{q}}(\psi^{a}J_{a}-{1\over 2}f_{ab}^{c}:\psi^{a}\psi^{b}\psi_{c}:),  
\nmb
G^{-}={2\over \sqrt{q}}(\psi_{a}J^{a}-{1\over 2}f^{ab}_{c}:\psi_{a}\psi_{b}\psi^{c}:),   
\nmb
K=(\delta^{b}_{a}+{2\ov q}Q^{b}_{a}):\psi^{a}\psi_{b}:+
          {2\over q}(\phi_{C}J^{C}-\phi^{C}J_{C})             
\nmb
T={1\ov q}:(J_{a}J^{a}+J^{a}J_{a}):+{2\ov q}(f_{\mu a}^{b}J^{\mu}-f^{\mu b}_{a}J_{\mu}):\psi^{a}\psi_{b}:+
\nmb
{1\ov 2q}(f_{\mu a}^{c}f^{d\mu}_{b}-f_{\mu b}^{c}f^{d\mu}_{a}-f_{\mu a}^{d}f^{c\mu}_{b}+f_{\mu b}^{d}f^{c\mu}_{a})
:\psi^{a}\psi^{b}\psi_{c}\psi_{d}:+
\nmb
{1\ov 2}(\dlt^{b}_{a}+{2\ov q}(f_{\mu n}^{b}f^{\mu n}_{a}+f_{\mu a}^{n}f^{\mu b}_{n})):(\d \psi^{a}\psi_{b}-\psi^{a}\d\psi_{b}):
\label{D.12}
\enr
where
\ber
Q^{b}_{a}=\phi_{M}f^{Mb}_{a}+\phi^{M}f_{Ma}^{b},
\nmb
\phi_{M}=f_{Mb}^{b}, \ \phi^{M}=f^{Mb}_{b}
\label{D.13}
\enr
generate the $N=2$ Virasoro
superalgebra (\ref{D.Vir}) with the central charge
\ber
 c=3(d+{2\ov q}Q^{a}_{a}) . 
\label{D.14}
\enr

 The statement of the theorem is more or less obvious if one notices that $G^{\pm}(z)$ are the BRST currents of $t_{\pm}$ subalgebras so the nilpotence of the currents follows from the OPE's for the currents $J^{a}(z)$ and $J_{a}(z)$ correspondingly.
To check the remaining $N=2$ superalgebra Virasoro OPE's one can just follow the analysis of Kazama and Suzuki. 

 It make sense however to prove the theorem by the direct calculation of the operator product expansions of the currents (\ref{D.12}) and using some important identities the structure constants of Manin triple $(g,g_{+},g_{-})$ and Manin subtriple $(h,h_{+},h_{-})$ satisfy according to the assumption. First of all we note that
\ber
f^{\mu a}_{\nu}=f_{\mu a}^{\nu}=0
\label{D.15}
\enr
because of $t_{\pm}$ are isotropic subalgebras and the form $(,)$ is invariant and nondegenerate. 
Due to (\ref{D.7}), (\ref{D.8}) and (\ref{D.15}) we obtain
\ber
G^{+}(z_{1})G^{-}(z_{2})=z_{12}^{-3}(2d-{2\ov q}<E^{a},E_{a}>-{2\ov q}f_{ab}^{c}f^{ab}_{c})+
\nmb
z_{12}^{-2}((2\dlt^{b}_{a}-{2\ov q}<E^{b},E_{a}>+{2\ov q}(2f_{ac}^{d}f^{cb}_{d}+f_{cd}^{b}f^{cd}_{a})):\psi^{a}\psi_{b}:(z_{2})+
{4\ov q}(\phi_{M}J^{M}-\phi^{M}J_{M})(z_{2}))+
\nmb
z_{12}^{-1}({2\ov q}(:J_{a}J^{a}:+:J^{a}J_{a}:)(z_{2})+{2\ov q}(\phi_{M}\d J^{M}-\phi^{M}\d J_{M})(z_{2})+
\nmb
{4\ov q}(f_{\mu a}^{b}J^{\mu}-f^{\mu b}_{a}J_{\mu}):\psi^{a}\psi_{b}:(z_{2})+
\nmb
{1\ov q}(f_{ab}^{n}f^{cd}_{n}-f_{an}^{c}f^{nd}_{b}+f_{bn}^{c}f^{nd}_{a}+f_{an}^{c}f^{dn}_{b}-f_{an}^{d}f^{cn}_{b})
:\psi^{a}\psi^{b}\psi_{c}\psi_{d}:(z_{2})+
\nmb
(2\dlt^{b}_{a}-{2\ov q}<E^{b},E_{a}>+{4\ov q}f_{ac}^{d}f^{cb}_{d})\d\psi^{a}\psi_{b}+{2\ov q}f_{cd}^{b}f^{cd}_{a}\psi^{a}\d\psi_{b}(z_{2}))+reg..
\label{D.16}
\enr
Using (\ref{D.4}) and (\ref{D.15}) we find that
\ber
-<E^{b},E_{a}>+2f_{ac}^{d}f^{cb}_{d}+f_{cd}^{b}f^{cd}_{a}=
2(f^{\mu n}_{a}f_{\mu n}^{b}+f^{b\mu}_{n}f_{a\mu}^{n}+f_{nm}^{b}f^{nm}_{a})
\label{D.17}
\enr
Then one can show using the last identity from (\ref{D.2}) that
\ber
f^{\mu n}_{a}f_{\mu n}^{b}+f^{b\mu}_{n}f_{a\mu}^{n}+f_{nm}^{b}f^{nm}_{a}=\phi_{M}f^{Mb}_{a}+\phi^{M}f_{Ma}^{b}=Q^{b}_{a}
\label{D.18} 
\enr
Similarly we find
\ber
f_{ab}^{n}f^{cd}_{n}-f_{an}^{c}f^{nd}_{b}+f_{bn}^{c}f^{nd}_{a}+f_{bn}^{d}f^{cn}_{a}-f_{an}^{d}f^{cn}_{b}=
\nmb
f_{\mu a}^{c}f^{d\mu}_{b}-f_{\mu b}^{c}f^{d\mu}_{a}-f_{\mu a}^{d}f^{c\mu}_{b}+f_{\mu b}^{d}f^{c\mu}_{a}
\label{D.19}
\enr
It gives
\ber
G^{+}(z_{1})G^{-}(z_{2})=z_{12}^{-3}2(d+{2\ov q}Q^{a}_{a})+
\nmb
z_{12}^{-2}2((\delta^{b}_{a}+{2\ov q}Q^{b}_{a}):\psi^{a}\psi_{b}:+
          {2\over q}(\phi_{C}J^{C}-\phi^{C}J_{C}))(z_{2})+
\nmb
z_{12}^{-1}({2\ov q}:(J_{a}J^{a}+J^{a}J_{a}):+{4\ov q}(f_{\mu a}^{b}J^{\mu}-f^{\mu b}_{a}J_{\mu}):\psi^{a}\psi_{b}:+
\nmb
{1\ov q}(f_{\mu a}^{c}f^{d\mu}_{b}-f_{\mu b}^{c}f^{d\mu}_{a}-f_{\mu a}^{d}f^{c\mu}_{b}+f_{\mu b}^{d}f^{c\mu}_{a})
:\psi^{a}\psi^{b}\psi_{c}\psi_{d}:+
\nmb
(\dlt^{b}_{a}+{2\ov q}(f_{\mu n}^{b}f^{\mu n}_{a}+f_{\mu a}^{n}f^{\mu b}_{n})):(\d \psi^{a}\psi_{b}-\psi^{a}\d\psi_{b}):+
\nmb
(\delta^{b}_{a}+{2\ov q}Q^{b}_{a})\d(:\psi^{a}\psi_{b}:)+
{2\over q}(\phi_{C}\d J^{C}-\phi^{C}\d J_{C}))+reg..
\label{D.20}
\enr
Thus we identify the central charge $c$, $U(1)$ current $K(z)$ and stress-energy tensor $T(z)$ according to (\ref{D.12}),
(\ref{D.14}). 

 Now we calculate the OPE
\ber
{\sqrt{q}\ov 2}K(z_{1})G^{+}(z_{2})=
z_{12}^{-2}\frac{2}{q}(Q^{n}_{k}f_{nm}^{k}+{1\ov 2}<\phi^{M}E_{M}-\phi_{M}E^{M},E_{n}>)\psi^{m}(z_{2})+
\nmb
z_{12}^{-1}{2\ov q}(Q^{b}_{a}\psi^{a}J_{b}-(\phi_{M}f^{MB}_{a}+\phi^{M}f_{Ma}^{B})\psi^{a}J_{B}+\phi_{M}f_{aB}^{M}\psi^{a}J^{B}+
\nmb
({1\ov 2}Q^{c}_{n}f_{ab}^{n}-Q^{n}_{a}f_{nb}^{c})\psi^{a}\psi^{b}\psi_{c})(z_{2})+
\nmb
z_{12}^{-1}{\sqrt{q}\ov 2}G^{+}(z_{2})+reg.
\label{D.21}
\enr

 It can be shown that 
\ber
\phi_{M}f_{aN}^{M}=0, \ \phi^{M}f^{aN}_{M}=0
\label{D.22a}
\enr
\ber
\phi_{M}f^{\mu M}_{a}=0, \ \phi^{M}f_{\mu M}^{a}=0
\label{D.22b}
\enr
because of (\ref{D.2}) and (\ref{D.15}). Hence,
\ber
{\sqrt{q}\ov 2}K(z_{1})G^{+}(z_{2})=
z_{12}^{-2}\frac{2}{q}(Q^{n}_{k}f_{nm}^{k}+{1\ov 2}<\phi^{M}E_{M}-\phi_{M}E^{M},E_{n}>)\psi^{m}(z_{2})+
\nmb
z_{12}^{-1}{2\ov q}
({1\ov 2}Q^{c}_{n}f_{ab}^{n}-Q^{n}_{a}f_{nb}^{c})\psi^{a}\psi^{b}\psi_{c}(z_{2})+
\nmb
z_{12}^{-1}{\sqrt{q}\ov 2}G^{+}(z_{2})+reg.
\label{D.23}
\enr
Because of Jacobi identity for $t_{+}$-subalgebra we obtain
\ber
Q^{c}_{n}f_{ab}^{n}-Q^{n}_{a}f_{nb}^{c}+Q^{n}_{b}f_{na}^{c}=\phi_{M}(f^{Mc}_{n}f_{ab}^{n}-f^{Mn}_{a}f_{nb}^{c}+f^{Mn}_{b}f_{na}^{c})
\label{D.24}
\enr
But the right hand side of this expression is vanishing due to (\ref{D.2}) and the first identity from (\ref{D.22b}) so that the only unwanted term in the right hand side of (\ref{D.21}) is the second order pole contribution. This contribution is vanishing also because
\ber
{1\ov 2}<\phi^{M}E_{M}-\phi_{M}E^{M},E_{m}>=\phi^{M}f_{Mk}^{n}f_{mn}^{k}+\phi_{M}f^{Mn}_{k}f_{mn}^{k}=-Q^{n}_{k}f_{nm}^{k}
\label{D.25}
\enr
where the (\ref{D.4}), the first identity from (\ref{D.22b}) and the definition of $Q^{a}_{b}$ have been used. Thus we obtain
\ber
K(z_{1})G^{+}(z_{2})=z_{12}^{-1}G^{+}(z_{2})+reg.
\label{D.26}
\enr
It can be shown similarly that the remaining OPE's from (\ref{D.Vir}) are satisfied also. It finishes the proof.
 
 Notice that because of $t_{\pm}$ are subalgebras and due to (\ref{D.15}) the subgroup $H$ of the Lie algebra $h$ is a Poisson-Lie subgroup of the Poisson-Lie group $G$ of the Lie algebra $g$. It makes the homogeneous space $G/H$ Poisson homogeneous space according to Semenov-Tian-Shansky \cite{SemTian} and Drinfeld \cite{Drin1} (see also \cite{LuW}). However, the Poisson structure on $G/H$ is a special one \cite{Drin1} so it would be interesting to see if one can generalize the Manin triple construction of N=2 superconformal coset models to include more general Poisson homogeneous spaces described in \cite{Drin1}, \cite{Karol}.

 It would be interesting to see if the Manin triple construction above can be extended to the N=4 superconformal coset models. The Manin triple construction of N=4 WZNW models has been represented in \cite{P2}.

\vskip 10pt
\centerline{\bf ACKNOWLEDGEMENTS}
\frenchspacing
I thank to A.Belavin and M.Bershtein for interest and discussions. The research was performed under a grant funded by Russian Science Foundation (project
No. 14-12-01383).

\end{document}